\documentclass[pre,aps,twocolumn]{revtex4-1}
\usepackage[english]{babel}

\usepackage{graphicx}
\usepackage{hyperref}

\begin{document}

\addtolength{\textheight}{-2.5cm}

\hspace{-0.5\textwidth}

\vspace*{21.6cm}
\hspace{-0.545\textwidth}
\noindent
\fbox{\begin{minipage}{1.0\textwidth}
\small\sf This is the accepted version of the following article: J.M. Escalante and S.E. Skipetrov, Longitudinal optical fields in light scattering from dielectric spheres and Anderson localization of light, \textit{Ann. Phys. (Berlin)} \textbf{529}, 1700039 (2017), which has been published in final form at \href{https://doi.org/10.1002/andp.201700039}{\tt https://doi.org/10.1002/andp.201700039}. This article may be used for non-commercial purposes in accordance with the \href{https://authorservices.wiley.com/author-resources/Journal-Authors/licensing/self-archiving.html}{Wiley Self-Archiving Policy}.
\end{minipage}}

\vspace{0.5cm}

\vspace*{-24cm}

\title{Longitudinal optical fields in light scattering from dielectric spheres\\ and Anderson localization of light}
\author{Jose M. Escalante}
\affiliation{Universit\'{e} Grenoble Alpes, LPMMC, F-38000 Grenoble, France}
\affiliation{CNRS, LPMMC, F-38000 Grenoble, France}
\author{Sergey E. Skipetrov}
\email{sergey.skipetrov@lpmmc.cnrs.fr}
\affiliation{Universit\'{e} Grenoble Alpes, LPMMC, F-38000 Grenoble, France}
\affiliation{CNRS, LPMMC, F-38000 Grenoble, France}
\begin{abstract}
Recent research has shown that coupling between  point scatterers in a disordered medium  by longitudinal electromagnetic fields  is   harmful for  Anderson  localization of light. However, it has been unclear if this feature is generic or specific for point scatterers. The present  work demonstrates that the intensity of longitudinal field outside a spherical dielectric scatterer illuminated by monochromatic light exhibits a complicated, nonmonotonous dependence on the scatterer size. Moreover, the intensity is reduced for a hollow sphere, whereas one can adjust the parameters of a coated sphere to obtain a relatively low longitudinal field together with a strong resonant scattering efficiency. Therefore, random arrangements of structured (hollow or coated) spheres  may be  promising three-dimensional disordered materials for reaching Anderson localization of light.
\end{abstract}
\maketitle

\section{Introduction}


Light scattered by a small dielectric particle exhibits a complicated spatial pattern of electromagnetic field featuring both transverse and longitudinal components. The latter are defined with respect to the vector $\mathbf{r}$ going from the particle center to the observation point. The longitudinal component of the electric field $E_r(\mathbf{r})$ is parallel to $\mathbf{r}$ and decays rapidly with the distance to the particle surface becoming negligible at distances of a few wavelengths $\lambda_0$. Therefore, it plays no role in light scattering from isolated particles (where standard measurements are performed in the far field) or in multiple light scattering from weakly disordered optical materials where typical distances between scattering events well exceed $\lambda_0$. When the latter condition is violated, near-field coupling between neighboring scatterers---including the coupling via the longitudinal field---yields corrections to the scattering properties of particle ensembles \cite{mcneil01,rezvani15}. More strikingly, longitudinal fields have been recently predicted to prevent Anderson localization of light---the complete halt of light transport due to strong disorder \cite{anderson58,john84,anderson85,lagendijk09}---in dense three-dimensional (3D) ensembles of point-like scatterers \cite{skip14}. Even in two dimensions, longitudinal fields may suppress localization when the electric field of the electromagnetic wave is in the plane of propagation \cite{maximo15}, in contradiction with the common believe that `all waves are localized in 2D' \cite{abrahams79,sheng06}.

The negative impact of longitudinal fields on Anderson localization is due to a nonradiative channel of energy transport that they provide. In dense molecular systems, such a transport mechanism is known since long time ago and carries the name of F\"{o}rster resonance energy transfer (FRET) \cite{forster48,forster59}; the discovery of this phenomenon was actually published by Theodor F\"{o}rster in this journal almost 70 years ago \cite{forster48}. Physically, FRET is due to the interaction between molecules via quasistatic dipole-dipole coupling when the inter-molecular distances are small enough. This interaction can be described in different `languages' as a classical dipole-dipole interaction or as an exchange of virtual photons (see, e.g., Ref.\ \cite{andrews04} for a recent discussion). The impact of nonradiative energy transfer due to the resonant dipole-dipole interaction and associated longitudinal electromagnetic fields on light propagation in disordered atomic systems has been considered by Nieuwenhuizen \emph{et al.} in the limit of low atomic densities \cite{nieu94}. The recent work \cite{skip14} demonstrated that for high atomic densities, this energy transfer mechanism becomes dominant and precludes Anderson localization of light in random 3D ensembles of point-like scatterers (atoms). Even though it is still unclear whether this result can be extended to other disordered media and, in particular, to random ensembles of dielectric particles of size $a \sim \lambda_0$, it seems that longitudinal optical fields arise as a new and previously overlooked aspect of such a long-standing problem of condensed-matter physics as Anderson localization. They may constitute a crucial obstacle on the way towards light localization by 3D disorder \cite{skip16} and may provide a plausible explanation for  the failure of attempts to observe Anderson localization of light in experiments \cite{wiersma97,beek12,storzer06,sperling13,sperling16}.

\addtolength{\textheight}{2.5cm}

In the present work, we report results of a theoretical study of the longitudinal optical field near a spherical scatterer illuminated by a monochromatic plane wave. For typical experimental situations (homogeneous, coated and hollow spheres made of TiO$_2$, silicon or GaAs, and silica), we calculate the longitudinal field intensity $I = |E_r|^2$ as well as the scattering efficiency $Q_s = \sigma_s/(\pi a^2)$  as functions of the size parameter of the scatterer $k_0 a$ (where $k_0 = 2\pi/\lambda_0$ is the wave number outside the scatterer, $a$ is the scatterer radius, and $\sigma_s$ is the scattering crosssection). As we discussed above, longitudinal-field coupling between scattering particles forming an optically dense disordered material may be harmful for Anderson localization whereas, on the other hand, localization can be reached only for strong scattering and hence large $Q_s$ \cite{sheng06}. We therefore explore the possibility of minimizing $I$ outside the scatterer near Mie resonances where $Q_s$ is resonantly enhanced. Our results yield an indication of scatterer parameters that may maximize chances of reaching Anderson localization of light in 3D, provided that the conclusion about the detrimental role of the longitudinal field made for point scatterers ($k_0 a \ll 1$) \cite{skip14,maximo15,skip15} remains valid for large scatterers ($k_0 a \gtrsim 1$) as well.

\section{Homogeneous sphere}
\label{homo}

We consider a homogeneous spherical particle (radius $a$, real refractive index $n$) located at the origin of a coordinate system and surrounded by vacuum or air (refractive index 1). The particle is illuminated by a linearly polarized along $x$ axis plane wave of unit amplitude (frequency $\omega$,  wave vector $\mathbf{k}_0 = k_0 \mathbf{e}_z$).
To quantify the magnitude of the longitudinal field $E_r(\mathbf{r})$ outside the scatterer, we use two parameters: the maximum intensity $I_{\mathrm{max}}$ and the normalized integral of intensity $W$ defined in the following way:
\begin{eqnarray}
&&I_{\mathrm{max}} = \max\limits_{r > a} \left| E_r(\mathbf{r}) \right|^2\,,
\label{imax}
\\
&&W = \frac{1}{a^3} \int\limits_{r>a} d^3 \mathbf{r} \left| E_r(\mathbf{r}) \right|^2\,.
\label{w}
\end{eqnarray}

\begin{figure*}
\centering{
\includegraphics[width=0.9\textwidth]{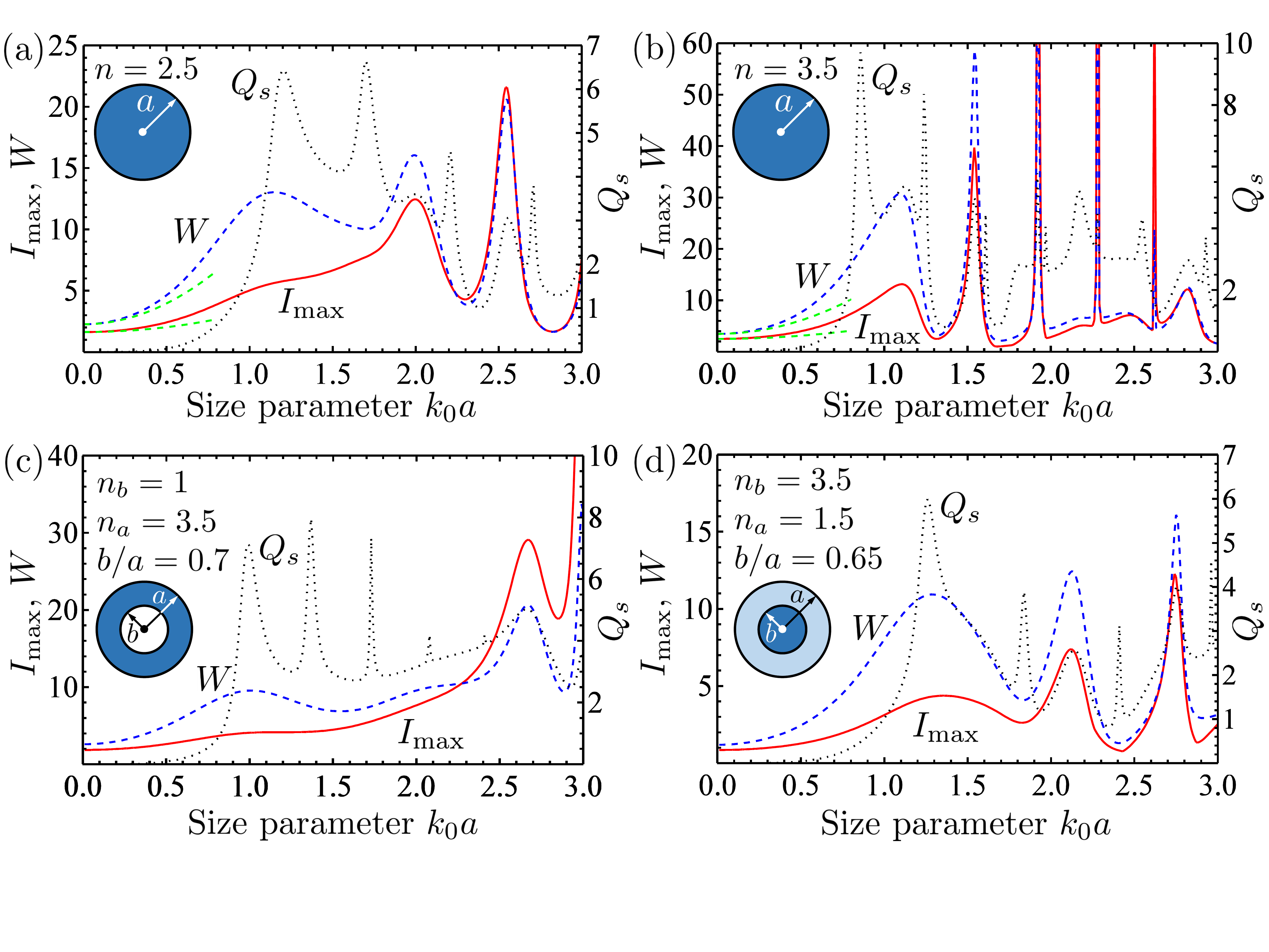}
}
\vspace{-1.5cm}
\caption{\label{fig1}The maximum intensity of longitudinal field $I_{\mathrm{max}}$ (red solid line), the integral of the longitudinal field intensity $W$ (blue dashed line), and the scattering efficiency $Q_s$ (black dotted line) for homogeneous (a,b), hollow (c), and coated (d) spheres. Green dotted lines show analytically calculated behaviors of $I_{\mathrm{max}}$ and $W$ for $k_0 a \ll 1$. $k_0$ is the wavenumber of the incident wave, $a$ is the sphere radius.}
\end{figure*}

For a small particle ($k_0 a \ll 1$), $I_{\mathrm{max}}$ and $W$ can be readily calculated. Indeed, the electric field $\mathbf{E}_0$ of the light incident on the sphere induces a dipole moment $\mathbf{p} = \alpha \mathbf{E}_0$, where $\alpha = 4 \pi \epsilon_0 a^3 (n^2 - 1)/(n^2 + 2)$ is the polarizability of the sphere and $\epsilon_0$ is the vacuum permittivity \cite{born99}. The oscillating dipole moment $\mathbf{p} = \mathbf{p}_0 \exp(-i \omega t)$ radiates a field \cite{born99}
\begin{eqnarray}
\label{emission}
\mathbf{E}(\mathbf{r}, t) &=& \frac{1}{4 \pi \epsilon_0} \left[
\frac{k_0^2}{r} (\hat{\mathbf{r}} \times \mathbf{p}_0) \times \hat{\mathbf{r}} \right.
\nonumber \\
&+& \left. \left( \frac{1}{r^3} - \frac{i k_0}{r^2} \right)
\left(3 \hat{\mathbf{r}} (\hat{\mathbf{r}} \cdot \mathbf{p}_0) - \mathbf{p}_0 \right) \right]
\nonumber \\
&\times& \exp(i k_0 r - i \omega t)\,,
\end{eqnarray}
of which the longitudinal component (i.e. the component parallel to $\mathbf{r}$) is
\begin{eqnarray}
\label{long}
E_r(\mathbf{r}, t) &=& \frac{1}{2 \pi \epsilon_0} \left( \frac{1}{r^3} - \frac{i k_0}{r^2} \right)
\left(\hat{\mathbf{r}} \cdot \mathbf{p}_0 \right)
\nonumber \\
&\times& \exp(i k_0 r - i \omega t)\,,
\end{eqnarray}
with $\hat{\mathbf{r}} = \mathbf{r}/r$. Using the proportionality between $\mathbf{p}$ and $\mathbf{E_0}$ and putting $E_0 = 1$ we readily obtain the intensity of the longitudinal field $I(\mathbf{r}) = \left| E_r(\mathbf{r}, t) \right|^2$:
\begin{eqnarray}
\label{intensity}
I(\mathbf{r}) &=& \left|
\frac{k_0^3 \alpha}{2 \pi \epsilon_0}
\left[ \frac{1}{(k_0 r)^3} - \frac{i}{(k_0 r)^2} \right]
\sin\theta \, \cos\varphi \right|^2\,.
\end{eqnarray}
The quantities defined by Eqs.\ (\ref{imax}) and (\ref{w}) are easily calculated from Eq.\ (\ref{intensity}):
\begin{eqnarray}
\label{imax2}
&&I_{\mathrm{max}} = 4 \left( \frac{n^2-1}{n^2+2} \right)^2 \left[ 1 + (k_0 a)^2 \right]\,,
\\
\label{w2}
&&W =  \frac{16 \pi}{9} \left( \frac{n^2-1}{n^2+2} \right)^2 \left[ 1 + 3(k_0 a)^2 \right]\,.
\end{eqnarray}
These results are shown by green dashed lines in Figs.\ \ref{fig1}(a) and (b).

For $k_0 a \gtrsim 1$ we need to go beyond the small-sphere approximation and use the Mie theory  \cite{bohren04}. Using the now standard notation of Ref.\ \cite{bohren04}, the longitudinal component of the scattered field can be written as
\begin{eqnarray}
E_r(\mathbf{r}, t) &=& \frac{i E_0 \cos(\varphi)}{k_0 r}
\sum\limits_{l = 1}^{\infty} i^l (2l + 1) a_l h_l^{(1)}(k_0 r)
\nonumber \\
&\times& P_l^1(\cos\theta) \exp(i k_0 r - i \omega t)\,,
\label{longmie}
\end{eqnarray}
where $h_l^{(1)}$ are the spherical Hankel functions of the first kind, $P_l^1$ are the associated Legendre polynomials, and $a_l$ are the scattering coefficients of Mie theory given by Eq.\ (\ref{al}) of Appendix A.

It follows from Eq.\ (\ref{longmie}) that the maximum of $I(\mathbf{r}) = \left| E_r(\mathbf{r}, t) \right|^2$ is reached at the surface of the sphere, i.e. at $r = a$, and for $\varphi = 0$ or $\pi$. We thus can write
\begin{eqnarray}
I_{\mathrm{max}} &=& \max\limits_{\theta} \left| \frac{1}{k_0 a} \sum\limits_{l = 1}^{\infty} i^l (2l + 1) a_l h_l^{(1)}(k_0 a) P_l^1(\cos\theta) \right|^2 \hspace{-2mm}.\;\;\;\;
\label{imaxhomo}
\end{eqnarray}
On the other hand, the integration of $I(\mathbf{r})$ over $\mathbf{r}$ yields
\begin{eqnarray}
W = 2 \pi \sum\limits_{l = 1}^{\infty} l(l + 1)(2l + 1) |a_l|^2 f_l(k_0 a)\,,
\label{whomo}
\end{eqnarray}
where
\begin{eqnarray}
f_l(k_0a) = \frac{1}{(k_0a)^3} \int\limits_{k_0a}^{\infty} dx \left| h_l^{(1)}(x) \right|^2\,.
\label{flka}
\end{eqnarray}

We present $I_{\mathrm{max}}$ and $W$ calculated from Eqs.\ (\ref{imaxhomo}) and (\ref{whomo}) in Figs.\ \ref{fig1}(a) and (b) as functions of the size parameter $k_0 a$ for two typical values of the particle refractive index corresponding to the two materials used in the experiments aimed at reaching the Anderson localization of light in 3D: $n = 2.5$ (TiO$_2$ in anatase phase \cite{storzer06,sperling13}) and $n = 3.5$ (GaAs \cite{wiersma97}). Figures\ \ref{fig1}(a) and (b) also show the scattering efficiency \cite{bohren04}
\begin{eqnarray}
Q_s = \frac{2}{(k_0a)^2} \sum\limits_{l=1}^{\infty} (2l + 1)
\left( |a_l|^2 + |b_l|^2 \right)\,,
\label{qshomo}
\end{eqnarray}
which exhibits sharp Mie resonances.
Calculation of $Q_s$ requires a second Mie coefficient $b_l$ defined by Eq.\ (\ref{bl}).

\begin{figure*}
\includegraphics[width=\textwidth]{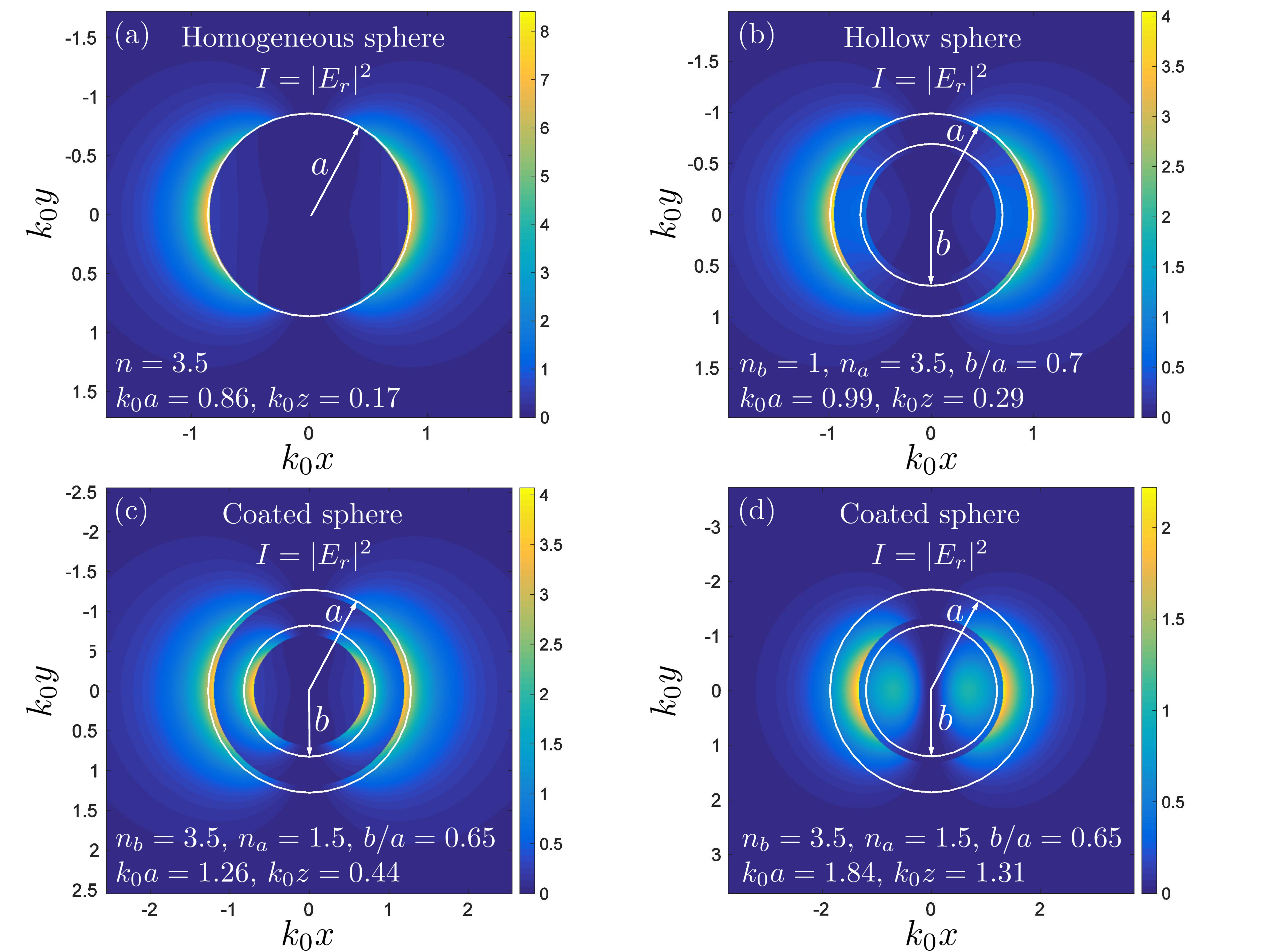}
\caption{\label{fig2}
False color plots of the intensity of longitudinal field component $I = |E_r|^2$ in the plane $z = \mathrm{const}$ for homogeneous (a), hollow (b) and coated (c,d) dielectric spheres. The panels (a--c) correspond to the first maxima of $Q_s$ in Fig.\ \ref{fig1}(b--d), respectively; the panel (d) corresponds to the second maximum of $Q_s$ in Fig.\ \ref{fig1}(d). For each panel, $z$ is chosen to correspond to the plane in which the maximum value of $I$ is reached. White circles show the boundaries of the sphere and of the core (for the hollow and coated spheres). This figure was realized using \textit{MatScat} software \cite{soft1,schafer02}.
}
\end{figure*}

Inspection of Figs.\ \ref{fig1}(a) and (b) shows that the variations of $I_{\mathrm{max}}$ and $W$ with $k_0 a$ do not follow the resonant structure of $Q_s$. Although both $I_{\mathrm{max}}$ and $W$ exhibit maxima and minima for certain values of $k_0 a$, the positions of these extrema do not generally coincide with resonances in $Q_s$. This signals that the properties of the longitudinal component of the scattered field are not simple functions of far-field quantities (such as, e.g., $Q_s$) and that it is therefore important to analyze them separately as we do in this work. On the other hand, $I_{\mathrm{max}}$ and $W$  have a tendency to grow and decrease together, especially for $k_0 a > 1$ where they closely follow each other. This validates the choice of these parameters as reliable measures of the longitudinal field strength. An important conclusion that follows from Figs.\ \ref{fig1}(a) and (b) is that in the range of $k_0 a$ corresponding to several first Mie resonances in $Q_s$, both $I_{\mathrm{max}}$ and $W$ take considerable values comparable with or exceeding those in the point-scatterer limit $k_0 a \to 0$. Therefore, the destructive impact of the longitudinal electromagnetic field on the phenomenon of Anderson localization predicted in the point-scatterer limit \cite{skip14}, is likely to persist for dielectric scatterers of size $a \sim \lambda_0$, which are typically used in experiments \cite{wiersma97,beek12,storzer06,sperling13,sperling16}\footnote{Even though we consider only spherical scatterers here, we expect this conclusion to be valid for scatterers of any reasonable shape.}.
On the other hand, one can see from Figs.\ \ref{fig1}(a) and (b) that the value of $k_0 a$ can be chosen to compromise between the largest possible $Q_s$ and the weakest possible $I_{\mathrm{max}}$ and $W$. In Fig.\ \ref{fig1}(b), for example, the third resonance in $Q_s$ at $k_0 a \simeq 1.25$ is close to the first minimum of $I_{\mathrm{max}}$ and $W$ at $k_0 a \simeq 1.3$. Therefore, if $k_0 a$ is adjusted somewhere in the range 1.25--1.3, one can expect to have both a relatively high scattering efficiency and a relatively weak longitudinal field.

A map of the longitudinal field intensity $I(\mathbf{r})$ in a plane $k_0 z = 0.17$, where $I(\mathbf{r})$ reaches its maximum value $I_{\mathrm{max}}$, is shown in Fig.\ \ref{fig2}(a) for $n = 3.5$ and $k_0 a = 0.86$ corresponding to the first and the strongest Mie resonance in the scattering efficiency $Q_s$ [see Fig.\ \ref{fig1}(b)]. We see that the maximum $I_{\mathrm{max}}$ of $I(\mathbf{r})$ is reached in the direction of the $x$ axis, i.e. in the direction that is parallel to the direction of oscillation of the induced dipole moment of the sphere, at least for a small sphere [see also Eq.\ (\ref{intensity}) that reaches its maximum for $\varphi = 0$ or $\pi$]. This is in contrast to the far-field radiation pattern of an oscillating dipole that exhibits a minimum in the direction of its oscillation ($x$ axis in our case).

\newpage
\section{Hollow and coated spheres}
\label{coated}

Some degree of control over the longitudinal electric field outside the scatterer can be achieved by using a spherical scatterer with internal structure. The simplest example of such a scatterer is a coated sphere---a sphere of radius $b$ and refractive index $n_b$ surrounded by a spherical shell of outer radius $a$ (shell thickness $a-b$) and refractive index $n_a$. A particular case of coated sphere is a hollow sphere for which $n_b = 1$. Light scattering by coated and hollow spheres has been extensively studied both theoretically \cite{bohren04,tinker72,pecora74,lock12,lock12a} and experimentally \cite{retsch11,ruck16,lasue07}. Synthesis of monodisperse sub-$\mu$m hollow TiO$_2$ spheres for potential use in photonic applications was demonstrated in by Eiden-Assmann \emph{et al.} \cite{eiden04}. Multiple light scattering experiments in concentrated suspensions of TiO$_2$ spheres coated with silica have been recently performed by Jimenez-Villar \emph{et al.} \cite{jimenez16} who have claimed observation of Anderson localization of light in these systems.

The calculation of the radial component of the electric field scattered by a hollow or coated sphere can be performed along the same lines as the one for a homogeneous sphere. We arrive at the same expressions (\ref{imaxhomo}) and (\ref{whomo}) for $I_{\mathrm{max}}$ and $W$ with a modified definition (\ref{alcoated}) of $a_l$.
The scattering efficiency of the coated sphere is given by Eq.\ (\ref{qshomo}) with $b_l$ defined by Eq.\ (\ref{blcoated}) \cite{bohren04}.

Results corresponding to hollow spheres are presented in Figs.\ \ref{fig1}(c) and \ref{fig2}(b) for a shell thickness equal to 30\% of the particle radius and the shell refractive index $n = 3.5$ (Si or GaAs). The comparison of Figs.\ \ref{fig1}(b) and (c) corresponding to homogeneous and hollow spheres with the same refractive index, shows that hollow spheres may have an advantage as compared to homogenous ones because they generate weaker longitudinal fields for the range of parameters corresponding to several first Mie resonances in $Q_s$. This advantage is, however, partially counterbalanced by the reduced scattering efficiency of the hollow sphere (by $\sim 25$\% at the first Mie resonance) with respect to a homogeneous one. Because we consider only a single, isolated scatterer, we cannot determine the extent to which suppressing the longitudinal field at the expense of reducing scattering efficiency may be beneficial for reaching Anderson localization of light in a large ensemble of identical scatterers. This would require an analysis of ensembles of at least several scatterers, which is beyond the scope of the present work.

A coated sphere with the refractive index $n_b$ of the core inferior to the refractive index $n_a$ of the coat exhibits a behavior that is intermediate between the homogeneous and the hollow sphere [Figs.\ \ref{fig1}(b) and (c), respectively]. The results for a coated sphere with $n_a < n_b$ are presented in Figs.\ \ref{fig1}(d) and \ref{fig2}(c,d). We have chosen $n_b = 3.5$ (Si or GaAs) and $n_a = 1.5$ (silica). In this case, varying the ratio $b/a$ at fixed $n_a$, $n_b$ allows shifting the resonant maxima of $Q_s$ and the minima of $I_{\mathrm{max}}$, $W$ with respect to each other, which may be an interesting property when one seeks to maximize scattering together with minimizing the longitudinal-field effects in an experiment. In Fig.\ \ref{fig1}(d), for example, we have adjusted $b/a$ to achieve a coincidence of the second and forth resonances of $Q_s$ with local minima of $I_{\mathrm{max}}$ and $W$. The comparison of spatial intensity maps of the longitudinal field [see Figs.\ \ref{fig2}(c) and (d)] shows that the field outside the scatterer is weaker at the second resonance of $Q_s$ [Fig.\ \ref{fig2}(d)] than at the first one [Fig.\ \ref{fig2}(c)]. However, similarly to the case of the hollow sphere, it is difficult to say whether this suppression of the longitudinal field, which is likely to be an advantage for reaching Anderson localization of light, is sufficient to compensate for the loss of scattering efficiency ($-30\%$) at the second resonance with respect to the first one. Analysis of light scattering by ensembles of several spheres is required to answer this question. Nevertheless, the possibility of controlling the relative locations of maxima in $Q_s$ and minima in $I_{\mathrm{max}}$, $W$ by varying the parameters of a coated dielectric sphere and, in particular, the possibility to achieve their coincidence, makes coated spheres interesting and potentially promising elementary scattering units of a disordered material in which Anderson localization of light may be eventually reached.

\section{Discussion}

Anderson localization of light in 3D---the main motivation for this work---cannot be realized in dilute ensembles of spherical scatterers where standard transport theory applies \cite{sheng06}. However, optical properties of ensembles of identical particles (which count for Anderson localization) can be deduced from those of a single particle (which we analyze in this paper) only in dilute media. Let us discuss how this limitation restricts application of our results to realistic experimental situations. First, we assumed that a dielectric particle is illuminated by a plane wave whereas a particle in a disordered medium composed of many particles, experiences waves scattered by other particles, which are not plane waves and which arrive from different directions. Although this seems to be an important aspect at a first sight, it is actually not, as far as all the waves have random, uncorrelated phases and do not interfere. Indeed, any arbitrarily complex wavefield can be decomposed in plane waves which then do not `feel' each other and interact with the particle as if there were no other waves. This is due to the linearity of the problem and the superposition principle, and holds at any particle number density $\rho$. Because particles considered in this work are spherically symmetric, parameters that minimize $I_{\mathrm{max}}$ or $W$ for a given incident plane wave will do so for all other incident waves as well. A problem may arise, however, if plane waves incident from different directions have correlated phases leading to a constructive interference of longitudinal components of scattered wavefields at a particular location. Here we neglect this possibility because we consider it statistically negligible.

A more serious issue concerns the scattering efficiency $Q_s$ and crossection $\sigma_s = Q_s \times \pi a^2$ of a single particle, which determine the scattering strength of the disordered material composed of many identical particles only at small $\rho$ because the scattering mean free path of a photon is $\ell = 1/\rho \sigma_s$ in this case\footnote{The scattering strength is also characterized by the transport mean free path $\ell^*$ that takes into account the anisotropy of scattering \cite{sheng06}, but it is sufficient to consider $\ell$ for our purposes. All our reasonings can be repeated for $\ell^*$ as well.}. In a dense medium (i.e., for large $\rho$), this simple relation between $\ell$ and $\sigma_s$ breaks down and the calculation of $\ell$ becomes much more involved even for point-like scatterers \cite{bvt90,bvt94,cherroret16}. The large size $a \sim \lambda_0$ of spheres considered in this work leads to additional complications due to inevitable correlations in sphere positions at large $\rho$ \cite{fraden90}. It may happen then that the maximum value of $\sigma_s$ or $Q_s$ does not correspond to the maximum of the scattering strength of the disordered material, which is quantified by $1/\ell$. We believe, however, that this complication should not compromise the main result of our analysis---the possibility of controlling longitudinal-field effects by changing scatterer parameters---though it will certainly modify the values of parameters $k_0 a$, $b/a$ for which longitudinal-field effects are minimized at given $n_a$, $n_b$. Numerical analysis of clusters composed of at least 2 or 3 spheres needed to further clarify this issue falls beyond the scope of this work.

Keeping in mind the above limitations of our analysis, let us now say a few words about the relation between our theoretical results and the existing experiments. The first claim of achieving Anderson localization of light in 3D was made for a GaAs powder with average particle diameter $2a = 300$ nm and light at a wavelength $\lambda_0 = 1064$ nm \cite{wiersma97}. The resulting size parameter $k_0 a \approx 0.9$ roughly corresponds to the first Mie resonance in Fig.\ \ref{fig1}(b) and hence to the maximum scattering efficiency. However, it follows from Fig.\ \ref{fig1}(b) that this value of $k_0 a$ is associated with a strong longitudinal field outside the scatterer, which  may lead to a strong coupling between nearby scatterers and prevent Anderson localization. This may explain the fact that later measurements did not confirm Anderson localization in the same and similar disordered media \cite{beek12}. The second claim of Anderson localization of light in 3D was based on light scattering experiments in powders of TiO$_2$ particles with diameters of several hundred nanometers \cite{storzer06,sperling13}. For one of the most strongly scattering samples $2a = 250$ nm and $k_0 a = 1.3$ at $\lambda_0 = 590$ nm. Again, according to Fig.\ \ref{fig1}(a) this corresponds to strong scattering but also to a strong longitudinal field outside each particle, which may be harmful for Anderson localization and may explain the absence of the latter established in the later work \cite{sperling16}. Finally, Jimenez-Villar \emph{et al.} have recently claimed Anderson localization of light in ethanol suspensions of TiO$_2$ particles coated by silica \cite{jimenez16}. Although our calculations indicate that using coated dielectric spheres may be potentially interesting for reaching Anderson localization of light, the irregular shape of scattering particles used in Ref.\ \cite{jimenez16} and their polydispersity wash out the minima that exist in the dependence of the longitudinal field intensity on the size parameter and do not allow us to quantify the role of longitudinal fields in these experiments.

\section{Conclusions}

We have studied the intensity of the longitudinal electric field near a dielectric spherical scatterer illuminated by a linearly polarized monochromatic plane wave. The intensity strongly depends on the size of the scatterer and on its internal structure. In particular, it is strongly suppressed at certain values of the size parameter $k_0 a$, which may be an advantage when one seeks to reach Anderson localization of light in a material comprised of a large number of identical spherical scatterers. For a homogeneous sphere, however, the values of $k_0 a$ minimizing the longitudinal field outside the sphere do not coincide with the values needed to benefit from Mie resonances and reach the highest scattering efficiency $Q_s$, even though the maxima of $Q_s$ can be close to minima of the longitudinal field intensity in the parameter space. A minimum of longitudinal field and a maximum of scattering efficiency may be achieved for the same values of parameters for a coated sphere. However, the scattering efficiency of a coated sphere is considerably reduced compared to a homogeneous sphere. Finally, a hollow sphere generates a longitudinal field that is weaker than that of a homogeneous sphere with the same refractive index, but it also has a lower scattering efficiency. Altogether our results indicate that suspensions or powders of structured (coated or hollow) spheres may be promising materials for continuing the search for Anderson localization of light in 3D. We hope that the use of structured scatterers may break the deadlock where this research field seems to get stalled after the works \cite{beek12,sperling16} that disproved the two experimental observations of Anderson localization in 3D available previously.

\begin{acknowledgements}
This work was funded by Agence Nationale de la Recherche (project ANR-14-CE26-0032 LOVE).
\end{acknowledgements}

\section*{Appendix A. Scattering coefficients of Mie theory}

In the main text, we use the following coefficients defined in the framework of Mie theory \cite{bohren04}:
\begin{eqnarray}
a_l &=& \frac{\psi_l(k_0a) \psi_l'(nk_0a) - n \psi_l'(k_0a) \psi_l(nk_0a)}{\xi_l(k_0a) \psi_l'(nk_0a) - n \xi_l'(k_0a) \psi_l(nk_0a)}\,,
\label{al} \\
b_l &=& \frac{n \psi_l(k_0a) \psi_l'(nk_0a) - \psi_l'(k_0a) \psi_l(nk_0a)}{n \xi_l(k_0a) \psi_l'(nk_0a) - \xi_l'(k_0a) \psi_l(nk_0a)}\,
\label{bl}
\end{eqnarray}
for the homogeneous sphere and
\begin{eqnarray}
a_l &=& \left( \psi_l(k_0a) [\psi_l'(n_ak_0a)-A_l \chi_l'(n_ak_0a)] \right.
\nonumber \\
&-& \left. n_a \psi_l'(k_0a) [\psi_l(n_ak_0a)-A_l \chi_l(n_ak_0a)] \right)/
\nonumber \\
&&\left( \xi_l(k_0a) [\psi_l'(n_ak_0a)-A_l \chi_l'(n_ak_0a)] \right.
\nonumber \\
&-& \left. n_a \xi_l'(k_0a) [\psi_l(nk_0a)-A_l \chi_l(n_ak_0a)]
\right)\,,
\label{alcoated}
\\
b_l &=& \left( n_a \psi_l(k_0a) [\psi_l'(n_ak_0a)-B_l \chi_l'(n_ak_0a)] \right.
\nonumber \\
&-& \left. \psi_l'(k_0a) [\psi_l(n_ak_0a)-B_l \chi_l(n_ak_0a)] \right)/
\nonumber \\
&&\left( n_a \xi_l(k_0a) [\psi_l'(n_ak_0a)-B_l \chi_l'(n_ak_0a)] \right.
\nonumber \\
&-& \left. \xi_l'(k_0a) [\psi_l(nk_0a)-B_l \chi_l(n_ak_0a)]
\right)
\label{blcoated}
\end{eqnarray}
for the coated and hollow spheres. The auxiliary coefficients $A_l$ and $B_l$ in Eqs.\ (\ref{alcoated}) and (\ref{blcoated}) are given by
\begin{eqnarray}
A_l &=& \frac{n_a \psi_l(n_ak_0b) \psi_l'(n_bk_0b) - n_b \psi_l'(n_ak_0b) \psi_l(n_bk_0b)}{n_a \chi_l(n_ak_0b) \psi_l'(n_bk_0b) - n_b \chi_l'(n_ak_0b) \psi_l(n_bk_0b)}\,, \nonumber \\
\label{aal} \\
B_l &=& \frac{n_a \psi_l(n_bk_0b) \psi_l'(n_ak_0b) - n_b \psi_l'(n_bk_0b) \psi_l(n_ak_0b)}{n_a \chi_l'(n_ak_0b) \psi_l(n_bk_0b) - n_b \psi_l'(n_bk_0b) \chi_l(n_ak_0b)}\,. \nonumber \\
\label{bbl}
\end{eqnarray}
The Riccati-Bessel functions are defined as
\begin{eqnarray}
\psi_l(x) &=& x j_l(x)\,, \\
\xi_l(x) &=& x h_l^{(1)}(x)\,, \\
\chi_l(x) &=& -x y_l(x)\,,
\end{eqnarray}
where $j_l(x)$ and $y_l(x)$ are the spherical Bessel function of the first and second kinds, respectively, and $h_l^{(1)}$ are the spherical Hankel functions of the first kind.

\end{document}